\documentclass[aps,prl,preprintnumbers,twocolumn,groupedaddress,nofootinbib]{revtex4}

\usepackage{latexsym}
\usepackage{amsmath}
\usepackage{amsfonts}
\usepackage{amsthm}
\usepackage{graphicx}

\usepackage{color}

\usepackage[latin1,utf8]{inputenc}
\usepackage[T1]{fontenc}

\allowdisplaybreaks

\newcommand{\bq}{\begin{eqnarray}}
\newcommand{\eq}{\end{eqnarray}}

\newcommand{\NF}{N_F}

\newcommand{\field}{\phi}
\newcommand{\coupling}{\lambda}

\theoremstyle{plain}

\newcommand{\arxivdate}{October 28, 2022}

\begin{document}

\preprint{MITP/22-090}
\title{\boldmath{Non-perturbative computation of lattice correlation functions by differential equations}}

\author{Federico Gasparotto, Andreas Rapakoulias and Stefan Weinzierl}
\affiliation{PRISMA Cluster of Excellence, Institut f{\"u}r Physik, Johannes Gutenberg-Universit\"at Mainz, D-55099 Mainz, Germany}

\date{\arxivdate}

\begin{abstract}
We show that methods developed in the context of perturbative calculations can be transferred to non-perturbative calculations.
We demonstrate that correlation functions on the lattice can be computed 
with the method of differential equations, supplemented with techniques from twisted cohomology.
We derive differential equations for the variation with the coupling or -- more generally -- with the parameters
of the action.
Already simple examples show that the differential equation with respect to the coupling 
has an essential singularity at zero coupling
and a regular singularity at infinite coupling.
The properties of the differential equation at zero coupling can be used to prove 
that the perturbative series is only an asymptotic series.
\end{abstract}

\maketitle

% -----------------------------------------------------------------------------
\section{Introduction}
\label{sect:intro}

Correlation functions (or scattering amplitudes in momentum space) are fundamental objects
in quantum field theory.
Standard methods to compute them are perturbation theory in the small coupling regime or
numerical lattice simulations in the strong coupling regime.
Not much is known for the analytic computation of correlation functions at finite coupling, some approaches in this direction have been considered in \cite{Schwarz:2008sa,Albert:2008ui,Gwilliam:2012jg,JohnsonFreyd:2012ww}.

Correlations functions usually require a regularisation scheme 
to regulate ultraviolet and/or infrared singularities.
In this letter we use lattice regularisation.
Lattice regularisation also implies that the space of all correlation functions is a finite-dimensional vector space \cite{Weinzierl:2020nhw}
and in this letter we exploit the finite-dimensionality of this vector space.

We show how to derive a first-order system of differential equations 
for the lattice correlation functions
with respect to the parameters appearing in the action (like the couplings).
Integration of the system of differential equations with appropriate boundary values gives the 
lattice correlation functions at finite coupling without resorting to perturbation theory.

Although the treatment in this letter is entirely non-perturbative, 
we transfer methods which are known from the computation of Feynman integrals 
to the computation of lattice integrals.
These are the well-established technique of integration-by-parts \cite{Tkachov:1981wb,Chetyrkin:1981qh},
the method of differential equations \cite{Kotikov:1990kg,Kotikov:1991pm,Remiddi:1997ny,Gehrmann:1999as}
and the formulation in terms of 
twisted cohomology \cite{aomoto1975,cho1995,Aomoto:book,Mizera:2017rqa,Mizera:2017cqs,Mastrolia:2018uzb,Frellesvig:2019kgj,Frellesvig:2019uqt,Mizera:2019vvs,Mizera:2019ose,Weinzierl:2020xyy,Frellesvig:2020qot,Caron-Huot:2021xqj,Caron-Huot:2021iev,Cacciatori:2021nli,Mastrolia:2022tww,Chestnov:2022alh,Chestnov:2022xsy,Giroux:2022wav}.
The following dictionary is useful:
The lattice integrals span a finite-dimensional vector space,
as do the Feynman integrals belonging to a specific topology.
Lattice integrals depend on parameters (the parameters appearing in the action, for example the coupling)
and so do Feynman integrals (they depend on kinematic variables).
In both cases we set up a differential equation, describing the variation with respect to these parameters.
In deriving this differential equation, integration-by-parts identities are used in both cases.
In a more mathematical language we have a vector bundle, where the parameters are coordinates on the base manifold
and the fibre is the above mentioned finite-dimensional vector space.
Instead of looking at integrals it is convenient to focus on the integrands, and here in particular on
equivalence classes of integrands with respect to a covariant derivative.
These equivalence classes are called twisted cocycles and both lattice integrals and Feynman integrals
can be discussed in this framework.

Our method works in principle for any space-time dimension and any lattice size.
However, with current algorithms the complexity grows exponentially with the number of lattice points 
and in practice one is limited to small lattices in low dimensions.
For these examples our method gives access to analytic properties of lattice correlation functions
not available from numerical Monte Carlo simulations.

Although the method to obtain the differential equations is similar between lattice correlation functions and
Feynman integrals, the resulting systems of differential equations are not: the known differential equations 
for Feynman integrals have only regular singularities.
This is not the case for the differential equations for lattice integrals: already in simple examples we encounter essential singularities.
This is expected in the continuum limit 
as it is believed that the perturbative series in the continuum limit is only an
asymptotic series (for a review see \cite{Suslov:2005zi}).
It is remarkable that already simple lattice systems show this behaviour:
We demonstrate that the perturbative series for a simple lattice example is only an asymptotic series.

% -----------------------------------------------------------------------------
\section{Notation}
\label{sect:notation}

We review the set-up from \cite{Weinzierl:2020nhw}.
We consider a lattice $\Lambda$ with lattice spacing $a$ in $D \in {\mathbb N}$ Euclidean space-time dimensions.
For simplicity we assume that the lattice consists of $L$ points in any direction. 
We assume periodic boundary conditions.
The lattice has $N=L^D$ points.
We label the lattice points by $x_1,\dots,x_N$ and 
denote the field at a lattice point $x$ by $\field_x$.
The field at the next lattice point in the (positive) $\mu$-direction modulo $L$ is denoted by
$\field_{x+a e_\mu}$.
We consider a scalar theory with Euclidean lattice action $S_E$ given by
\bq
\label{action}
 S_E
 =
 \sum\limits_{x \in \Lambda}
 \left( 
 -  \sum\limits_{\mu=0}^{D-1} \field_x \field_{x+a e_\mu}
 + D \field_x^2
 + \sum\limits_{j=2}^{j_{\mathrm{max}}} \frac{\coupling_j}{j!} \field_x^j
 \right),
\eq
with $\coupling_j \ge 0$ and $\coupling_{j_{\mathrm{max}}} \neq 0$.
We call $\coupling_{j_{\mathrm{max}}}$ the leading coupling.
Of particular interest are the cases $j_{\mathrm{max}} =3$, which is  called a $\phi^3$-theory 
and $j_{\mathrm{max}} =4$, which is called a $\phi^4$-theory.

We are interested in the lattice integrals
\bq
\label{lattice_integral}
 I_{\nu_1 \nu_2 \dots \nu_N}
 & = &
 \int\limits_{{\mathcal C}^N} d^N\field \left( \prod\limits_{k=1}^N \field_{x_k}^{\nu_k} \right) \exp\left(-S_E\right).
\eq
The integration contour ${\mathcal C}$ is a curve in ${\mathbb C}$
and the same for every field variable $\field_x$.
The integration contour is chosen such that $\exp(-S_E)$ goes to zero as we approach the boundary. 
The correlation functions are then given by
\bq
 G_{\nu_1 \nu_2 \dots \nu_N}
 & = &
 \frac{I_{\nu_1 \nu_2 \dots \nu_N}}{I_{0 0 \dots 0}}.
\eq 

% -----------------------------------------------------------------------------
\section{The differential equation}
\label{sect:method}

We may reformulate the lattice integrals in the language of twisted cocycles.
We define a function $u$, a one-form $\omega$
and a $N$-form $\Phi$ by
\bq
\label{def_twisted}
 u & = & \exp\left(-S_E\right),
 \nonumber \\
 \omega & = & d \ln u \;= \; -d S_E \; = \; \sum\limits_{x \in \Lambda} \omega_x d\field_x,
 \nonumber \\
 \Phi & = & \left( \prod\limits_{k=1}^N \field_{x_k}^{\nu_k} \right) d^N\field.
\eq
In terms of these quantities we may rewrite the integral in eq.~(\ref{lattice_integral}) as
\bq
 I_{\nu_1 \nu_2 \dots \nu_N}
 & = &
 \int\limits_{{\mathcal C}^N} u \; \Phi.
\eq
The one-form $\omega$ defines a covariant derivative $\nabla_\omega=d+\omega$.
By assumption,
the integrand vanishes on the boundary of the integration.
This leads to integration-by-parts identities.
In terms of $\Phi$ this translates to the statement that the integral is invariant under transformations
\bq
\label{equivalence_relation}
 \Phi' & = & \Phi + \nabla_\omega \Xi,
\eq
for any $(N-1)$-form $\Xi$. In addition, $\Phi$ is obviously $\nabla_\omega$-closed.
It is therefore natural to consider the twisted cohomology group $H_\omega^N$ defined 
as the equivalence classes of $\nabla_\omega$-closed $N$-forms modulo exact ones. We denote the equivalence classes by $\langle \Phi |$
and refer to these as twisted cocycles. In a similar way we denote the
integration cycle by $| {\mathcal C}^N \rangle$ and refer to it as a twisted cycle.
We also write
\bq
 I_{\nu_1 \nu_2 \dots \nu_N}
 & = &
 \left\langle \Phi \left| {\mathcal C}^N \right. \right\rangle
\eq
to emphasize that the integral is a pairing between a twisted cocycle and a twisted cycle.
In the following we focus on the twisted cocycles $\langle \Phi |$.
Our method relies on the fact that the twisted cohomology group $H^N_\omega$ is finite-dimensional.
For $\phi^{j_{\mathrm{max}}}$-theory the dimension is given by
\bq
 \NF & = & \dim H^N_\omega
 \; = \;
 \left(j_{\mathrm{max}}-1\right)^N.
\eq
A basis $\langle e_1 |, \dots, \langle e_{\NF} |$ is given by \cite{Weinzierl:2020nhw}
\bq
\label{def_basis}
 \left( \prod\limits_{k=1}^N \field_{x_k}^{\nu_k} \right) d^N\field,
 & &
 0  \; \le \; \nu_k \; \le \;  j_{\mathrm{max}}-2.
\eq
Using intersection numbers we may express any $\langle \Phi |$ as a linear combination of the basis $\langle e_i |$:
\bq
\label{reduction_to_basis}
 \left\langle \Phi \right| 
 & = &
 \sum\limits_{i=1}^{\NF} c_i \left\langle e_i \right|.
\eq
The coefficients $c_i$ are independent of the field variables $\field_x$.
Integrating both sides over the twisted cycle $|{\mathcal C}^N\rangle$ we see that 
we may write any lattice integral as
\bq
\label{reduction_to_basis_integral}
 \left\langle \Phi \left| {\mathcal C}^N \right. \right\rangle
 & = &
 \sum\limits_{i=1}^{\NF} c_i \left\langle e_i \left| {\mathcal C}^N \right. \right\rangle.
\eq
Let us denote by
\bq
 I_1 \; = \; \left\langle e_1 \left| {\mathcal C}^N \right. \right\rangle,
 & \ldots, &
 I_{\NF} \; = \; \left\langle e_{\NF} \left| {\mathcal C}^N \right. \right\rangle
\eq
the set of lattice integrals corresponding to the basis of twisted cocycles in eq.~(\ref{def_basis}).
This set spans the vector space of all lattice integrals. 
It does not need to be a basis: There could be additional (trivial) relations due to integration, not seen
at the level of the integrands (e.g. $dz_1 \neq dz_2$ but $\int\limits_0^1 dz_1 = \int\limits_0^1 dz_2$).

Let us now consider the derivative of $I_{\nu_1 \dots \nu_N}$ with respect to the coupling $\coupling_j$.
Taking the derivative of the exponential brings down extra factors of the field variables and we obtain
for the scalar theory of eq.~(\ref{action})
\bq
 \frac{d}{d\coupling_j} I_{\nu_1 \dots \nu_N}
 & = &
 - \frac{1}{j!} \sum\limits_{i=1}^N I_{\nu_1 \dots (\nu_i+j) \dots \nu_N}.
\eq
With the help of eq.~(\ref{reduction_to_basis_integral}) we may re-express the right-hand side 
as a linear combination of the spanning set $I_1, \dots, I_{\NF}$.
Doing this for every element $I_i$ of the spanning set yields
\bq
\label{general_dgl}
 \frac{d}{d\coupling_j} I_{i}
 & = &
 \sum\limits_{k=1}^{\NF}
 A_{ik} I_{k}.
\eq
This is the sought-after differential equation.
Readers familiar with the method of differential equations for Feynman integrals will certainly recognise the analogy.
We emphasise that the same technique can be applied to a non-perturbative problem.
The differential equation~(\ref{general_dgl}) and appropriate boundary values determine the correlation functions at all values of the coupling
and in particular at non-small values of the coupling.

% -----------------------------------------------------------------------------
\section{Example 1: $\phi^3$-theory}
\label{sect:example_1}

\begin{figure}
\begin{center}
\includegraphics[scale=1.0]{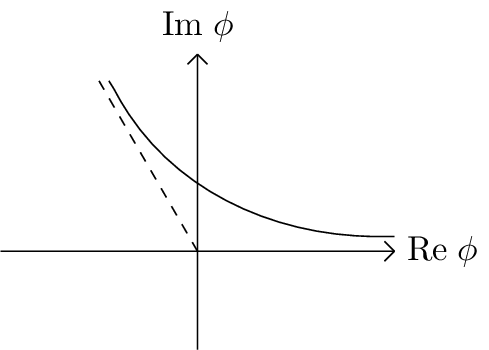}
\end{center}
\caption{
The integration contour for $\phi^3$-theory. 
The asymptotic values are $\arg \field = 2 \pi /3$ and $\arg \field = 0$.
}
\label{fig1}
\end{figure}

As our first example we consider massless $\phi^3$-theory in $D=1$ space-time dimensions with $L=2$ lattice points.
We set $\lambda_3=\kappa$.
The action is given by
\bq
 S_E
 & = &
 \left(\field_{x_1}-\field_{x_2}\right)^2 + \frac{\kappa}{6} \left(\field_{x_1}^3+\field_{x_2}^3\right).
\eq 
The potential in $\phi^3$-theory is not bounded from below for real values of the field variables.
In order to satisfy the condition that $\exp(-S_E)$ goes to zero as we approach the boundary, we take
as integration contour 
a contour with asymptotic values $\arg \field = 2 \pi /3$ and $\arg \field = 0$, 
as shown in fig.~\ref{fig1}.
The space of twisted cocycles is four dimensional. A possible spanning set of lattice integrals is given by
\bq
 I_{00}, I_{01}, I_{10}, I_{11}.
\eq
However, it is more convenient to work with the spanning set
\bq
 J_1 = I_{00},
 \;
 J_2 = I_{10}+I_{01},
 \;
 J_3 = I_{11},
 \;
 J_4 = I_{10}-I_{01},
 \;
\eq
as this spanning set will decouple the differential equation into a $(3\times3)$-block and a $1\times 1$-block.
We find the differential equations
\bq
\label{example_dgl_phi_3}
 \frac{d}{d\kappa}
 \left(\begin{array}{c} J_1 \\ J_2 \\ J_3 \\ \end{array} \right)
 & = &
 \left( \begin{array}{ccc}
 -\frac{2}{3\kappa} & 0 & -\frac{4}{3\kappa} \\
 \frac{8}{3\kappa^2} & -\frac{1}{\kappa} & \frac{32}{3\kappa^2} \\
 -\frac{32}{3\kappa^3} & \frac{4}{3\kappa^2} & -\frac{128}{3\kappa^3}-\frac{4}{3\kappa} \\
 \end{array} \right)
 \left(\begin{array}{c} J_1 \\ J_2 \\ J_3 \\ \end{array} \right),
 \nonumber \\
 \frac{d}{d\kappa}
 J_4
 & = &
 \left( -\frac{128}{3\kappa^3}-\frac{1}{\kappa} \right) J_4.
\eq
These differential equations have an essential singularity at $\kappa=0$ and a regular singularity at $\kappa=\infty$.
It is no surprise that the differential equations have a singularity at $\kappa=0$, as for $\kappa=0$ the factor $\exp(-S_E)$
no longer tends to zero as we approach the boundary of the integration contour.

The differential equations have the solution
\bq
 J_1
 & = &
 \frac{c_1}{\kappa^{2}} \; {}_2F_2\left(\frac{5}{6},\frac{7}{6};\frac{4}{3},\frac{5}{3};\frac{64}{3\kappa^2}\right)
 \nonumber \\
 & &
 +
 \frac{c_2}{\kappa^{\frac{4}{3}}} \; {}_2F_2\left(\frac{1}{2},\frac{5}{6};\frac{2}{3},\frac{4}{3};\frac{64}{3\kappa^2}\right)
 \nonumber \\
 & &
 +
 \frac{c_3}{\kappa^{\frac{2}{3}}} \; {}_2F_2\left(\frac{1}{6},\frac{1}{2};\frac{1}{3},\frac{2}{3};\frac{64}{3\kappa^2}\right),
 \nonumber \\
 J_4
 & = &
 \frac{c_4}{\kappa} \exp\left(\frac{64}{3\kappa^2}\right),
\eq
where ${}_2F_2$ denotes a hypergeometric function.
$c_1$-$c_4$ are four boundary constants.
The integrals $J_2$ and $J_3$ are obtained by differentiation of $J_1$:
\bq
 J_2
 & = &
 - \frac{9\kappa^3}{16} \frac{d^2 J_1}{d\kappa^2}
 - \left(24 + \frac{27\kappa^2}{16} \right) \frac{d J_1}{d\kappa}
 - \left(\frac{8}{\kappa} + \frac{\kappa}{2} \right) J_1,
 \nonumber \\
 J_3
 & = &
 - \frac{3\kappa}{4} \frac{d J_1}{d\kappa}
 - \frac{1}{2} J_1.
\eq
The lattice integrals have the symmetry
\bq
\label{symmetry_indices}
 I_{\nu_1 \nu_2} & = & I_{\nu_2 \nu_1},
\eq
which implies $J_4=0$ and $c_4=0$.
Note however that the differential equation is derived from the integrands.
The integrand of $I_{10}$ is not identical to the integrand of $I_{01}$.

It is possible to go to larger lattices and we give some indications for the required CPU time on a standard laptop:
For example, it takes about 
$190 \, \mathrm{s}$ to compute the differential equation for $D=1$ space-time dimensions with $L=10$ lattice points
and about $160 \, \mathrm{s}$ to compute the differential equation for $D=3$ space-time dimensions with $L=2$ lattice points
in each direction.

% -----------------------------------------------------------------------------
\section{Example 2: $\phi^4$-theory}
\label{sect:example_2}

As our second example we consider 
massive $\phi^4$-theory in $D=1$ space-time dimensions with $L=2$ lattice points.
We set $\lambda_2=m^2$ and $\lambda_4=\lambda$.
The action is given by
\bq
 S_E
 = 
 \left(\field_{x_1}-\field_{x_2}\right)^2
 + \frac{m^2}{2} \left(\field_{x_1}^2+\field_{x_2}^2\right)
 + \frac{\lambda}{24} \left(\field_{x_1}^4+\field_{x_2}^4\right).
 \nonumber
\eq 
As integration contour ${\mathcal C}$ we take the real axis.
The space of twisted cocycles is nine dimensional. A possible spanning set of lattice integrals is given by
\bq
 I_{00}, I_{01}, I_{02},
 I_{10}, I_{11}, I_{12},
 I_{20}, I_{21}, I_{22}.
\eq
Again, it is more convenient to work with the spanning set
\begin{align}
 J_1 & = I_{00},
 &
 J_2 & = I_{11},
 &
 J_3 & = I_{22},
 \nonumber \\
 J_4 & = I_{20}+I_{02},
 &
 J_5 & = I_{10}+I_{01},
 & 
 J_6 & = I_{21}+I_{12},
 \nonumber \\
 J_7 & = I_{10}-I_{01},
 & 
 J_8 & = I_{21}-I_{12},
 &
 J_9 & = I_{20}-I_{02}.
\end{align}
This decouples the differential equation with respect to $\lambda$ into a
$(4 \times 4)$-block consisting of $(J_1,J_2,J_3,J_4)$,
a $(2 \times 2)$-block consisting of $(J_5,J_6)$,
another $(2 \times 2)$-block consisting of $(J_7,J_8)$
and a $(1 \times 1)$-block consisting of $(J_9)$.
As before we have the symmetry of eq.~(\ref{symmetry_indices}), hence $J_7$, $J_8$ and $J_9$ are identical zero.
In addition we have that the integrands of $I_{10}$, $I_{01}$, $I_{21}$ and $I_{12}$ are antisymmetric under
$\field_{x_1} \rightarrow -\field_{x_1}, \field_{x_2} \rightarrow -\field_{x_2}$.
It follows that for the integration contour ${\mathcal C}$ along the real axis $J_5$ and $J_6$ are zero as well.
Hence, the interesting differential equation is the one for the integrals $(J_1,J_2,J_3,J_4)$.
This differential equation reads
\begin{widetext}
\bq
\label{dgl_system}
 \frac{d}{d\lambda}
 \left(\begin{array}{c}
 J_1 \\ J_2 \\ J_3 \\ J_4 \\
 \end{array} \right)
 & = &
 \left( \begin{array}{cccc}
 - \frac{1}{2\lambda} & - \frac{1}{\lambda} & 0 & \frac{2+m^2}{4 \lambda} \\
 0 & - \frac{3\left(2+m^2\right)^2}{\lambda^2} - \frac{1}{\lambda} & - \frac{1}{\lambda} & \frac{3\left(2+m^2\right)}{\lambda^2} \\
 - \frac{72}{\lambda^3} & - \frac{72\left(6+4m^2+m^4\right)}{\lambda^3} & - \frac{3\left(2+m^2\right)^2}{\lambda^2} - \frac{3}{2\lambda} & \frac{3 \left(2+m^2\right)\left(72+\lambda\right)}{2\lambda^3} \\
 \frac{3\left(2+m^2\right)}{\lambda^2} & \frac{18\left(2+m^2\right)}{\lambda^2} & \frac{2+m^2}{2\lambda} & - \frac{3\left(12+4m^2+m^4\right)}{2\lambda^2} - \frac{1}{\lambda} \\
 \end{array} \right)
 \left(\begin{array}{c}
 J_1 \\ J_2 \\ J_3 \\ J_4 \\
 \end{array} \right).
\eq
\end{widetext}
This differential equation has a regular singularity at $\lambda=\infty$ and
an essential singularity at $\lambda=0$.

We may re-write the system of four coupled first order differential equations as an ordinary fourth order differential
equation for one specific integral, say $J_1=I_{00}$:
\bq
\label{Picard_Fuchs}
 \left[
 \frac{d^4}{d\lambda^4}
 + p_3 \frac{d^3}{d\lambda^3} 
 + p_2 \frac{d^2}{d\lambda^2} 
 + p_1 \frac{d}{d\lambda} 
 + p_0
 \right] I_{00} & = & 0.
 \;
\eq
$p_3$-$p_0$ are rational functions of $\lambda$ and $m^2$ and given in an ancillary file attached to the arxiv version of this article.

We now investigate if this differential equation has a convergent series solution of the form
\bq
\label{power_series_solution}
 \lambda^\rho \sum\limits_{n=0}^\infty a_n \lambda^n.
\eq
We call such a solution a regular solution.
Let us denote by $o_j$ the order of the pole of $p_j$ at $\lambda=0$ and 
set $g_j=j+o_j$.
We find
\bq
 \left(g_3,g_2,g_1,g_0\right)
 & = &
 \left( 5, 6, 7, 6 \right).
\eq
As $g_1=7$ is the maximum among these numbers, it follows that there is at most one regular solution
around $\lambda=0$ (see \cite{Ince:book}, p.417).

Let us now assume that a solution of the form as in eq.~(\ref{power_series_solution}) exists.
From the indicial equation it follows that $\rho=0$.
From the differential equation in eq.~(\ref{Picard_Fuchs}) we obtain a recursion relation for the
coefficients $a_n$:
\bq
\label{recurrence_relation}
 a_n
 & = &
 r_1 a_{n-1} + r_2 a_{n-2} + r_3 a_{n-3} + r_4 a_{n-4}.
\eq
The $r_j$ are rational functions of $n$ and $m^2$ and given in an ancillary file attached to the arxiv version of this article.
The recursion relation determines together with the initial condition $a_n=0$ for $n<0$ all coefficients
in terms of one unknown coefficient $a_0$.
The function $r_4$ is given by
\bq
\lefteqn{
 r_4
 = - \frac{\left(n-2\right)\left(n-3\right)\left(2n-5\right)\left(2n-7\right)}{216 n}
 } & &
 \\
 & &
 \times
 \frac{\left(m^4+4m^2+12\right)}{m^6\left(m^2-2\right)\left(m^2+4\right)^3\left(m^2+6\right)\left(m^4+4m^2-4\right)}
\nonumber 
\eq
and is non-zero for $n\ge4$.
Together with the fact that $a_4\neq 0$ this shows that the series does not terminate.
The proof is simple: Assume that the series terminates and let $a_k$ be the last non-zero value. We then have $a_{k+4}=r_4 a_k$.
The left-hand side is zero due to our assumption, the right-hand side is non-zero.
This is a contradiction and our assumption is wrong.

As the series does not terminate it necessarily diverges for any non-zero value of $\lambda$ (see \cite{Ince:book}, p. 421).
The radius of convergence of the series in eq.~(\ref{power_series_solution}) is therefore zero, contradicting our assumption that a regular solution
around $\lambda=0$ exists.

The series 
\bq
\label{asymptotic_series}
 \sum\limits_{n=0}^\infty a_n \lambda^n
\eq
is an asymptotic series and coincides with the perturbative series.
For this simple example it is easy to compute the perturbative series as all integrals reduce to Gaussian integrals.
One finds
\bq
\label{generating_series}
\lefteqn{
 a_n
 = 
 \frac{2\pi}{\sqrt{m^2\left(4+m^2\right)}}
 \frac{\left(-1\right)^n}{n! 24^n}
 \sum\limits_{k=0}^n
 \left( \begin{array}{c} n \\ k \\ \end{array} \right)
 } & &
 \\
 & &
 \left.
 \left( \frac{\partial}{\partial S_1} \right)^{4k}
 \left( \frac{\partial}{\partial S_2} \right)^{4\left(n-k\right)}
 e^{\frac{2\left(S_1+S_2\right)^2+m^2\left(S_1^2+S_2^2\right)}{2m^2\left(4+m^2\right)}}
 \right|_{S_1=S_2=0}
 \nonumber
\eq
and in particular
\bq
 a_0
 & = &
 \frac{2\pi}{\sqrt{m^2\left(4+m^2\right)}}.
\eq
One may verify that the coefficients $a_n$ computed from eq.~(\ref{generating_series}) satisfy the recurrence relation of eq.~(\ref{recurrence_relation}).
Truncating an asymptotic series to the first few terms gives approximations, which improve with the number of included terms
up to a certain truncation order. Beyond this order the series will start to diverge. 
We have checked that this happens for $\lambda=1/2$ and $m=1$ around $n=15$ and for $\lambda=1/5$ and $m=1$ around $n=45$.

We may compare the solution of the differential equation with lattice Monte Carlo results.
\begin{figure}
\begin{center}
\includegraphics[scale=0.7]{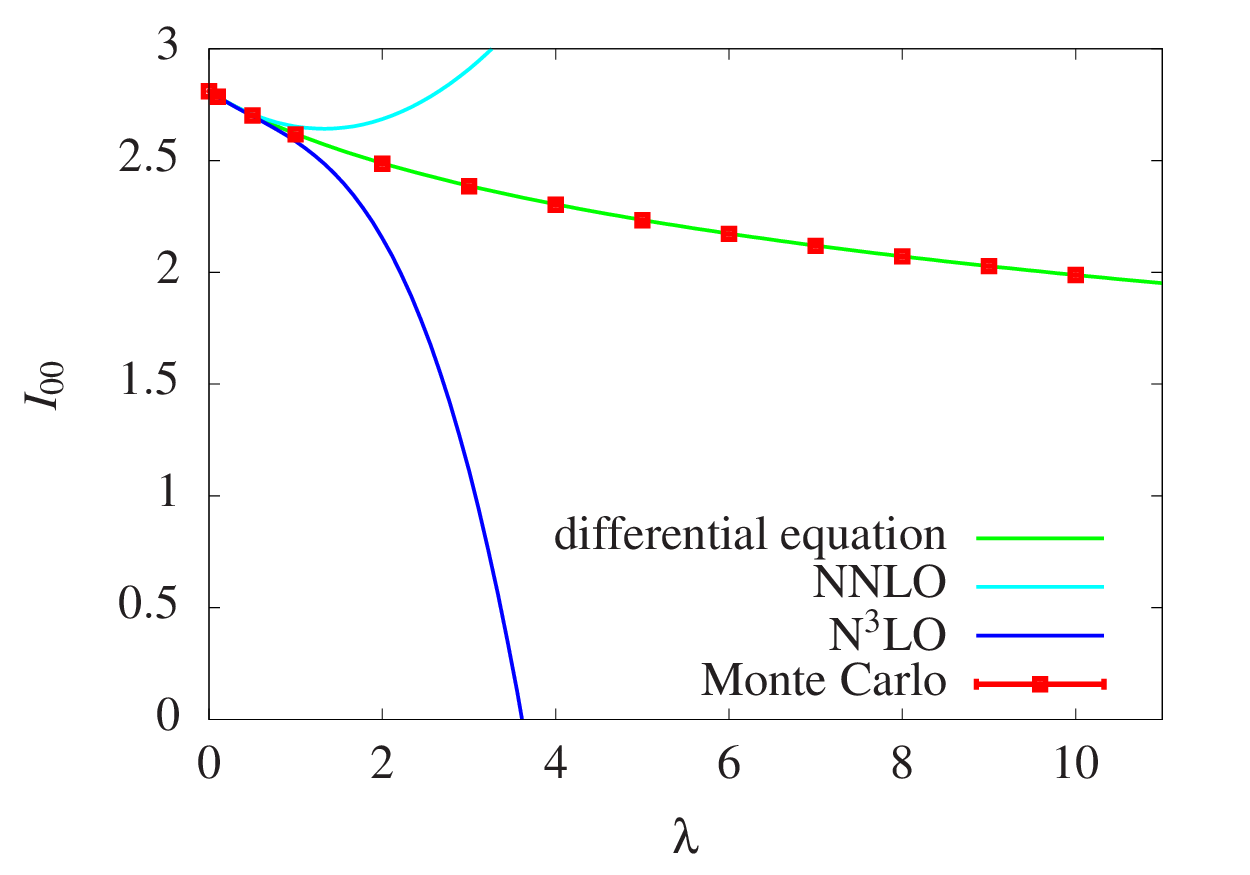}
\end{center}
\caption{
The lattice integral $I_{00}$ as a function of the coupling $\lambda$ obtained from the differential equation (green) and lattice Monte Carlo integration (red).
In addition, the plot shows the perturbative result truncated to third (cyan) and fourth (blue) order. 
}
\label{fig2}
\end{figure}
This is shown in fig.~\ref{fig2} for the lattice integral $I_{00}$ 
as a function of the coupling $\lambda$ for the fixed value of the mass parameter $m=1$.
For the solution of the differential equation we use boundary values at small coupling obtained from the perturbative series.
Fig.~\ref{fig2} shows that the solution of the differential equation and the lattice Monte Carlo results agree perfectly for all values of $\lambda$.
Fig.~\ref{fig2} also shows truncated perturbative series predictions: We plot the NNLO and N${}^3$LO predictions.
As expected, they agree reasonably at small values of the coupling, but differ significantly at non-small values of the coupling.

Let us summarise: The perturbative series of eq.~(\ref{asymptotic_series}) with the coefficients given by eq.~(\ref{generating_series})
is just an asymptotic series for $I_{00}$ around $\lambda=0$.
On the other hand, $I_{00}$ satisfies the fourth-order differential equation in eq.~(\ref{Picard_Fuchs}) (or equivalently the system in eq.~(\ref{dgl_system})).
The differential equation is more general, as it contains information on $I_{00}$ at any value of the coupling.

Also for $\phi^4$-theory we give some indications for the required CPU time on a standard laptop for larger lattices:
For example, it takes about 
$280 \, \mathrm{s}$ to compute the differential equation for $D=1$ space-time dimensions with $L=8$ lattice points
and about $1400 \, \mathrm{s}$ to compute the differential equation for $D=3$ space-time dimensions with $L=2$ lattice points
in each direction.

% -----------------------------------------------------------------------------
\section{Conclusions}
\label{sect:conclusions}

In this article we have shown how to obtain differential equations for lattice integrals, describing the
variation of these integrals with the parameters appearing in the action.
Supplemented with appropriate boundary values the differential equations determine the lattice integrals
for any values of the action parameters and thus go beyond perturbation theory.

Analysing the differential equation for a simple system already shows that the perturbative series for this system
is only an asymptotic series.

The dimension of the vector space of lattice integrals can be smaller than the dimension of the twisted cohomology group.
We observe that by an appropriate choice of basis for the twisted cohomology classes we may decouple the system of differential 
equations.

The techniques used to derive the differential equations are transferred from methods
developed in the context of perturbative calculations.
We expect that further refinements of these methods will be beneficial to both fields.

% -----------------------------------------------------------------------------
% references
\bibliography{/home/stefanw/notes/biblio}
\bibliographystyle{/home/stefanw/latex-style/h-physrev5}

\end{document}